\documentclass[aps,prb,reprint,onecolumn,superscriptaddress]{revtex4-2}
\usepackage{amsmath}
\usepackage{amssymb}
\usepackage{mathtools}
\PassOptionsToPackage{sort}{natbib}
\usepackage{natbib}
\usepackage{booktabs} 
\usepackage{float} 
\usepackage{placeins}
\usepackage{graphicx}
\usepackage{subcaption} 
\DeclareCaptionType[within=section]{mat}[Figure] 
\usepackage[margin=1in]{geometry}
\begin{document}
	\title{Absolutely Convergent Real-Space Madelung Summation Using Axial Multipoles}
	\author{Joven V. Calara} 
	\affiliation{Dept of Engineering, Salt Lake Community College}
	\email{jcalara@slcc.edu}
	\email{jvcalara@yahoo.com}
	\author{Jan D. Miller}
	\affiliation{Dept of Metallurgy, University of Utah}
	\email{jan.miller@utah.edu}
	\begin{abstract}
			We present an absolutely convergent real-space method for evaluating Madelung potentials in ionic lattices. The method, based on repeated axial multipole units with systematically eliminated low-order moments, applies uniformly to bulk crystals, surfaces, edges, interstitial sites, and exterior points, without recourse to reciprocal-space techniques. In the RU-13 construction, the far-field contribution of each axial multipole unit decays as $r^{-13}$, ensuring fast and absolute convergence of the real-space direct sum. The method yields identical limits under spherical and cubic summation geometries and reproduces standard Madelung constants with high precision, achieving ~13-digit accuracy within a radius of 40 lattice spacings. Extensions to dimensions $d=1–6$ exhibit convergence consistent with asymptotic decay predictions. 
	\end{abstract}
\maketitle
\section*{Introduction}
	Electrostatic or Coulombic forces account for a large share of the binding energies in ionic crystals, and therefore strongly influence their properties. But the long range properties of Coulombic forces make the seemingly straightforward calculation of total electrostatic potentials quite deceptive.\\\\
	In the interior of the lattice, the electrostatic potential energy of an ion is the sum of the potentials due to all the other ions in the crystal. For an infinite NaCl latttice as example, the potential energy U of an ion at the origin of a coordinate system parallel to the Bravais vectors is the sum of all contributions from the other ions;
\begin{equation}\label{Usum}
	U = \frac{e^2}{4\pi\varepsilon_0\alpha}\sideset{}{'}\sum_{ijk}^{\infty}\frac{(-1)^{i+j+k}}{\sqrt{i^2+j^2+k^2}}
\end{equation}
	where i,j,k are the integer coordinates of the alternating Na and Cl ions, $e$ is the electronic charge, $\epsilon_0$ is the permittivity of vacuum, and $\alpha$ a characteristic lattice parameter, often the nearest neighbor distance. The prime on the summation sign indicates self potential of the ion on the origin is excluded.\\\\
	The summation factor is identified as the Madelung constant. Following conventional practice for brevity, the pre-sum factor ${e^2}/{4\pi\varepsilon_0\alpha}$ will be dropped but is implied in expressions for potentials in the following discussions.\\\\
	The sum is conditionally convergent (i.e. it will diverge to $\pm\infty$ if all charges are of one sign). As such, it may converge to different limits, or not at all, depending on the order the terms are summed. The two general methods of attacking the convergence problems are direct summations and integral methods. Direct summation is the literal application of equation (\ref{Usum}), where potentials from individual or grouped charges are added as they are encountered by expanding volumes. A widely known direct summation is that of Evjen\cite{Evjen}, who grouped the NaCl charges into neutral shells of increasing size around a reference ion. The outermost assigned partial charges; 1/2, 1/4, and 1/8 if they are on a face, edge, or corner respectively. This eliminates the zeroth pole (charge), dipole, and quadrupole electrical moments, leaving the octupolar moment the dominant term. This renders the far-field potential decay at a distance $r$ as $r^{-4}$. This made the summation converge faster than naive summation. More recently Tavernier \cite{Tavernier} mapped the NaCl lattice into a Clifford torus, eliminating the effects of edge boundaries. This topological scheme enabled fast real-space convergence, but it is intrinsically tailored to bulk periodicity and is less directly applicable to broken-symmetry or off-lattice environments.\\\\ Of the integral methods, Ewald's\cite{Ewald} approach is the most widely known but involves integration in reciprocal space. This paper focuses only on real-space summation.\\\\
	The attraction of the real-space, or direct, summation is its apparent, but deceiving, simplicity. Evjen's method was a model advancement, but is dependent on identifying neutral shells around a reference ion, which is not always easy or possible. A natural approach is to add ions as they are encountered by expanding spherical shells as was done by Harrison\cite{Harrison} and Pratt\cite{Pratt}. The shells were found to be non-neutral and resulted in seemingly random swings above and below the target value. More relevantly, Borwein \cite{Borwein} revisited Emersleben’s proof that spherical summation is always divergent. Wolf \cite{Wolf} also used spherical truncation, but proposed to add thin spherical envelop enclosing the summation volume with a sufficient charge to preserve neutrality. The neutralizing sphere is similar to Harrison’s\cite{Harrison} proposal.  Other attempts use the expanding neutral, prismatic shells dictated by the lattice's Bravais vectors\cite{Essen}. This works well with NaCl and CsCl, but such shells are difficult to identify in the more complex structures such as the perovskite and rutile lattices.
	\\\\Suppression of the leading electrical moments has long been recognized as a route to faster real-space convergence. Baker \cite{Baker} noted that both quadrupole and lower moments must vanish, and that this condition is satisfied for CsCl by rhombohedral summation rather than by cubical growth along the Bravais vectors, which produces alternating pure Cs or Cl planes. Gelle\cite{Gelle} and Lepetit later formulated a general renormalized-cell construction for arbitrary crystal structures. Their method forms a supercell from translated unit cells and assigns weights to the charges so that prescribed multipole moments of the supercell cell vanish. The target lattice is then constructed by superposing translated copies of the supercell, thereby obtaining a rapidly converging summation.
	 \\\\The common theme in these previous works is that absolute convergence requires elimination of the lowest moments. The present work shows how axial multipoles can be designed to eliminate low-order moments by construction, yielding direct Madelung sums with algebraically controlled convergence.
	\section*{Axial Multipoles}
	Axial multipoles, as used here, are finite sets of discrete charges arranged on a common axis and chosen so that specified low-order electrical moments vanish. The exploratory use such charge arrays in Madelung summation was noted previously by the authors,\cite{Calara} in a brief communication. The present work gives the full construction, derives the moment-cancellation conditions, relates the first nonvanishing moment to the far-field decay, and shows how this decay controls the absolute convergence and numerical rate of the resulting real-space Madelung sums. 
	For a lattice built from identical neutral repeat units, absolute convergence follows if the far-field potential of each unit decays faster than \(R^{-3}\), where \(R\) denotes the discrete distance from the repeat unit to the reference ion at the origin. If the repeat-unit potential decays as \(R^{-p}\), its contribution to the lattice sum has the form
	
	\[
	\sum_{\mathbf{R}}{R^{-p}}.
	\]
	
	For large \(R\), the discrete lattice may be
	approximated by a continuum of points with spherical shell density
	\(4\pi r^2dr\), giving
	
	\[
	\sum_{\mathbf{R}} R^{-p}
	\;\sim\;
	\int^\infty 4\pi r^2 \frac{dr}{r^p}
	=
	4\pi\int^\infty r^{2-p}\,dr .
	\]
	The integral converges only for $p>3$. Hence, in three dimensions, the leading far-field term of each neutral repeating unit must decay faster than $r^{-3}$. Cancellation through the quadrupole moment leaves the octupolar term, whose potential decays as $r^{-4}$, and is therefore sufficient for absolute convergence.\\\\
	\begin{figure}[htbp]
		\centering
		\includegraphics[width=.7\linewidth]{"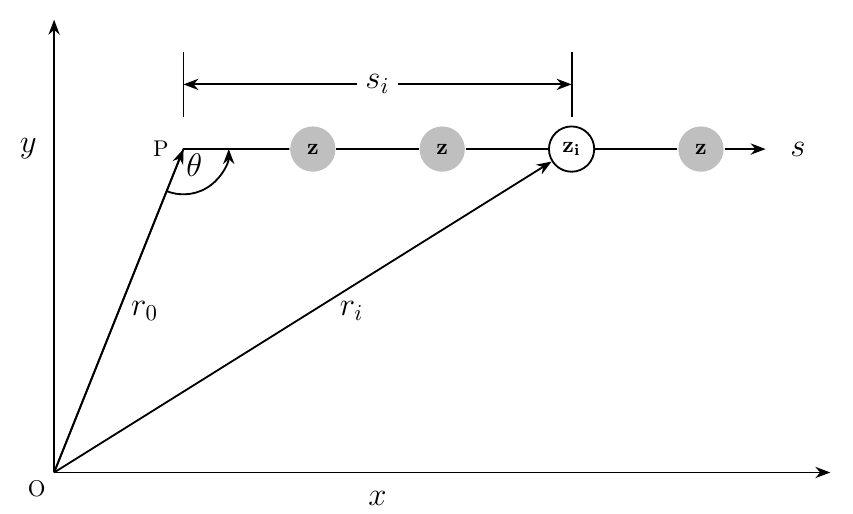"}
		\captionof{mat}{Charge array}
		\label{Array1}
	\end{figure}~\\
	Consider an array of point charges (Fig.~\ref{Array1}), $z_i, i=1,n,$ on a line S parallel to the x-axis. Point P on S is in the near neighborhood of the array but is otherwise arbitrarily located, and is at a distance $r_0$ from the origin O. Line OP makes an angle $\theta$ with line S. Charge $z_i$ is at a distance $s_i$ from P and at distance $r_i$ from the origin.\\ \\The contribution $\epsilon_i$ of the charge $z_i$ to the potential at the origin is;
	\begin{equation}
	\epsilon_i=\frac{z_i}{r_i}=\frac{z_i}{\sqrt{r_0^2+s_i^2-2r_0s_icos\theta}}
	=\frac{z_i}{r_0}\Bigr(1+\big(\frac{s_i}{r_0}\big)^2-2\frac{s_i}{r_0}cos\theta\Bigr)^{-\frac{1}{2}}
	\end{equation}
	Expanding the radical in a Taylor series and summing over the $n$ charges, the total potential $\epsilon_T$ from the array is then
\begin{equation}\label{eps-t}
	\epsilon_T=\sum_{i=1}^n\epsilon_i=
	\frac{P_0}{r_0}\sum_{i=1}^nz_i +
	\frac{P_1}{r_0^2}\sum_{i=1}^n{z_is_i} +
	\frac{P_2}{r_0^3}\sum_{i=1}^n{z_is_i^2} +
	\frac{P_3}{r_0^4}\sum_{i=1}^n{z_is_i^3} +
	\frac{P_4}{r_0^5}\sum_{i=1}^n{z_is_i^4}\ldots + \frac{P_k}{r_0^{k+1}}\sum_{i=1}^nz_is_i^k\ldots
\end{equation}
	where $P_i$s are Legendre polynomials (in $cos\theta)$(see, e.g.,\:Griffiths\cite{Griffiths}).\\
	\\
	The coefficient of each power of $(1/r_0)$ is the electric moment of order $k$, $\sum_i z_i s_i^k$, aside from the angular factor $P_k(\cos\theta)$. The contributions with $k=0,1,2,3,\ldots$ correspond respectively to the monopole, dipole, quadrupole, octupole, and higher moments, respectively.
	\\\\
	The series is ordered in powers of $s_i/r_0$. For $s_i<<r_0$, each term dominates over the next, so that the potential $\epsilon_T$ is determined by the first non-vanishing moment. \\\\
	%
%
\clearpage
\textbf{Construction of axial multipoles}\\\\
	Equation (\ref{eps-t}) shows that eliminating successive low-order moments makes the far-field potential decay progressively faster. Therefore, to make the collective potential decay as $r^{-5}$, for example, the first four terms in Eq.(\ref{eps-t}) must vanish identically. This requirement gives four simultaneous equations, after the common prefactors are removed:
\begin{equation}\label{simul01}
	\sum_{i=1}^nz_i = 0;\quad
	\sum_{i=1}^n{z_is_i} = 0;\quad
	\sum_{i=1}^n{z_is_i^2} = 0;\quad
	\sum_{i=1}^n{z_is_i^3} = 0
\end{equation}
At a far field $s_i<<r_0$, the the potential is equal to the first non-zero term;
\begin{equation}\label{epsRU5}
	\epsilon_T=
	\frac{P_4}{r_0^5}\sum_{i=1}^n{z_is_i^4},
\end{equation}
with the desired decay rate. \\\\
The four constraints of (\ref{simul01}) admit only the trivial solution when
$n \le 4$. The smallest nontrivial construction therefore requires
$n=5$, introducing one degree of freedom that must be fixed by an additional normalization condition.

To illustrate the procedure, consider the one-dimensional alternating
NaCl chain. Choosing the chain axis as the $x$-axis and placing five
charges $z_i$ at positions $x_i=i$ $(i=1,\ldots,5)$ with alternating
signs $(+-+-+)$, Eq.(\ref{simul01}) becomes:
\begin{equation}\label{simul02}
	\sum_{i=1}^{5} z_i x_i^m = 0,
	\qquad m=0,1,2,3.
\end{equation}

The fifth equation is obtained from charge normalization. Since the
positive-charge sublattice of NaCl consists of unit charges, the sum of
the positive coefficients must equal unity:
\begin{equation}\label{fifth}
	z_1 + z_3 + z_5 = 1
\end{equation}
Combining set of equations (\ref{simul02}) and (\ref{fifth}) yields;
\begin{equation}\label{mat01}
	\begin{bmatrix}
		1 & 1 & 1 & 1 & 1 \\
		1 & 2 & 3 & 4 & 5 \\
		1 & 2^2 & 3^2 & 4^2 & 5^2 \\
		1 & 2^3 & 3^3 & 4^3 & 5^3 \\
		1 & 0 & 1 & 0 & 1\\
	\end{bmatrix}
	\begin{bmatrix}
		z_1 \\
		z_2 \\
		z_3 \\
		z_4 \\
		z_5 \\
	\end{bmatrix}
=
	\begin{bmatrix}
		0 \\
		0 \\
		0 \\
		0 \\
		1
	\end{bmatrix}
\end{equation}
whose solution is
\begin{equation*}
	\begin{bmatrix}
		z_1 & z_2 & z_3 & z_4 & z_5
	\end{bmatrix}
=
	\begin{bmatrix}
		\dfrac{1}{8} & -\dfrac{1}{2} & \dfrac{3}{4} & -\dfrac{1}{2} & \dfrac{1}{8}
	\end{bmatrix}
\end{equation*}
\captionof{mat}{NaCl RU-5 axial hexadecapole. Decay rate is as \(r^{-5}\).}\label{RU-5}~

\begin{figure}[htbp]
	\centering
	\includegraphics[width=0.7\linewidth]{"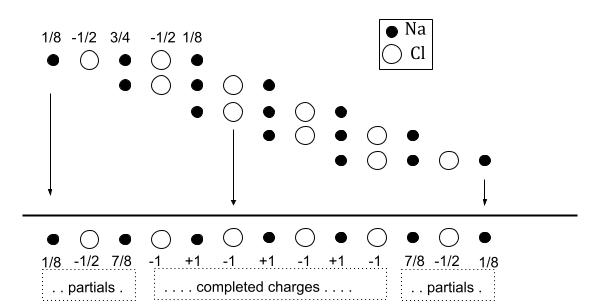"}
	\captionof{mat}{NaCl lattice being assembled from RU-5.}
	\label{NaCL_assembly}
\end{figure}

The resulting five-member repeating unit is shown in Fig.~2. Its first nonvanishing moment is the hexadecapole, and its far-field potential decays as \(r^{-5}\). We denote this unit RU-5. In the notation RU-\(p\), the suffix \(p\) identifies the leading asymptotic decay \(r^{-p}\), not necessarily the number of charges in the unit. For the present lattice, the two numbers happen to coincide.\\\\
The RU-5 construction translates easily to higher decay order designs. To obtain a unit whose far-field term term is proportional to $r^{-p}$, all lower-order moments up to $(p-2)$ must vanish. If we use the same charge positioning, as previously, on integer locations $x_i=i$ $(i=1,\ldots,N)$, the charges $z_i$ are required to satisfy
\[
\sum_{i=1}^{N}z_ix_i^k=0 ,\quad k=0,1,2,3 \ldots (p-2)
\]
where $N$ is the number of charges. For NaCl-like lattice, we choose $N=p$, the minimum for a non-trivial solution.
%
%
We then form the normalization equation that sums the positive charges to unity; $\sum(z_i>0)=1$.
The resulting linear system, analogous to Eq.\:(\ref{mat01}), yields the charges of the repeating unit. Table I shows the structures of several RUs. Note that the numerators of the partial charges are the coefficients of the binomial expansion of \({(1-x)^{p-1}}\).
\begin{table}[htbp]
	\centering
	\caption{Charge structure of selected RU-$p$ units. The suffix \textit{p} correspond to a potential decay as $\boldsymbol{r^{-p}}$. The common divisors of the fractional charges are listed separately in the \textit{Norm-factor} column for clarity. \textit{(For $p\!>\!13$ round off errors begin to affect accuracy).}}
	\begin{tabular}{cclllllllllllcc}
		\toprule[0.2em]
		Unit & \multicolumn{14}{c}{Structure}  \\
		\cmidrule{2-15}
		RU-p & Norm-factor \quad & \(z_1\)  & \(z_2\) & \(z_3\)& \(z_4\)& \(z_5\)& \(z_6\)& \(z_7\)& \(z_8\)& \(z_9\)& \(z_{10}\)& \(z_{11}\)& \(z_{12}\)& \(z_{13}\)\\
		\midrule
		RU-5     & 1/8  & 1 & -4 & 6 & -4 & 1 &&&&&&&& \\
		RU-6    & 1/16  & 1 & -5 & 10 & -10 & 5 & -1 &&&&&&&\\
		RU-7    & 1/32  & 1 & -6 & 15 & -20 & 15 & -6 & 1 &&&&&&\\
		RU-9    & 1/128  & 1 & -8 & 28 & -56  & 70  & -56  & 28  & -8  & 1  &   &&&  \\
		RU-13    & 1/2048  & 1 & -12 & 66 & -220 & 495 & -792 & 924 & -792 & 495 & -220 & 66 & -12 & 1\\
		\bottomrule[0.1em]
	\end{tabular}%
	\label{tab:RU-n}%
\end{table}\\\\
It is straightforward to verify that the RUs reproduce the NaCl lattice when adjacent units overlap with matching charge signs. Figure (\ref{NaCL_assembly}) illustrates how RU-5 units assemble to reproduce a line of NaCl sublattice.The terminals ends always consist of the same set of fractional charges, whereas the interior segment of unit charges lengthens. Different RUs will produce distinct end structures.
\section*{Madelung Constants}
\begin{figure}[htbp]
	\centering
	\includegraphics[width=0.6\linewidth]{"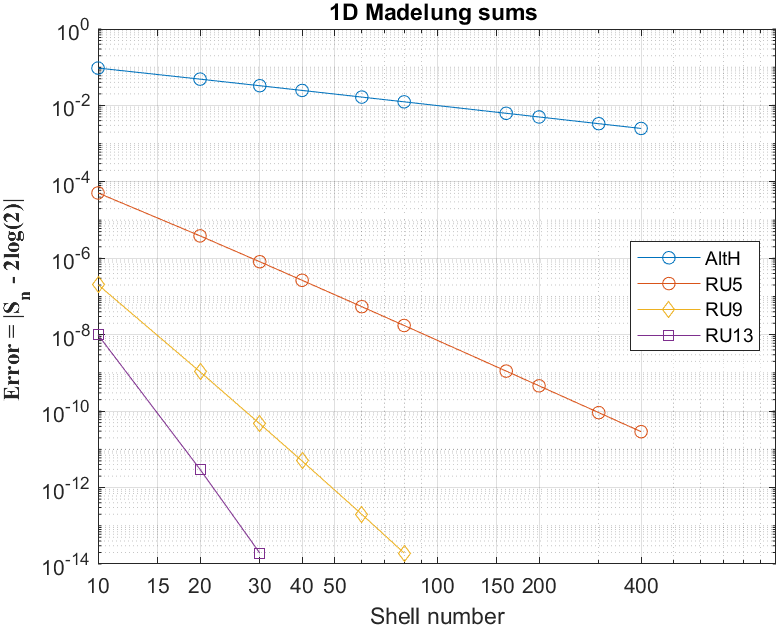"}
	\captionof{mat}{Convergence speeds of AltH, RU-5, RU-9, and RU-13 constructions over 1D NaCl lattice. $S_n$ is the partial sum to $nth$ shell.}
	\label{1D_speeds}
\end{figure}

\begin{figure}[htbp]
	\centering
	\includegraphics[width=0.7\linewidth]{"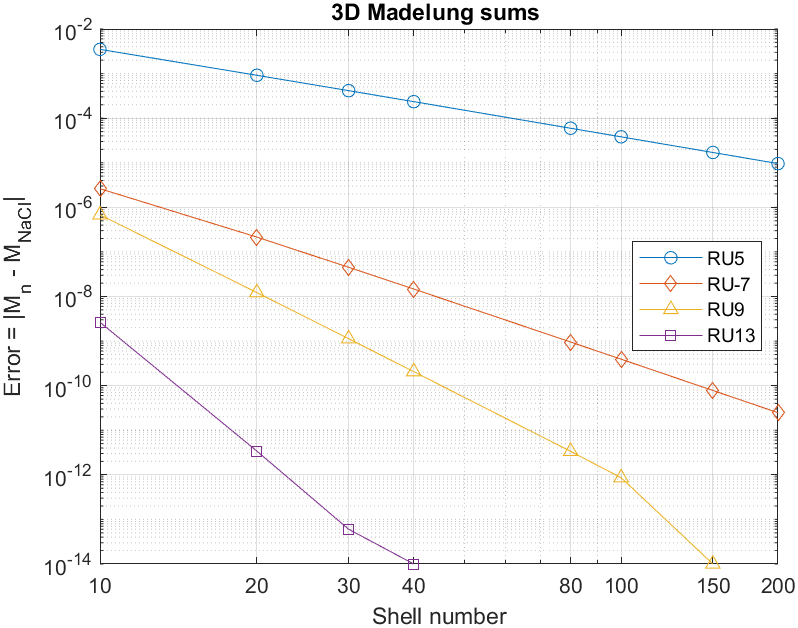"}
	\captionof{mat}{Convergence speeds of RU-5, RU-7, RU-9, and RU-13 constructions over 3D NaCl lattice. $M_n$ is the partial sum at the $n^{th}$ shell. NaCl bulk Madelung constant $M_{NaCl}$ = 1.7475645946331821 (Burrows\cite{Burrows}).}
	\label{3D_speeds}
\end{figure}

~\\
\textit{\textbf{\underline{1D NaCl}}}~\\\\
For an infinite one-dimensional NaCl lattice with alternating charges located at integer coordinates, with Na at the origin, the Madelung constant is
\begin{equation}\label{AltH}
	\epsilon_{1D}=2*(1-1/2+1/3-1/4+1/5-1/6\ldots)
\end{equation}
The factor 2 comes from the 2-fold symmetry of the lattice. In the parentheses of Eq. (\ref{AltH}) is alternating harmonic series for $\log_e(2)$, which gives 
\begin{equation}
	\epsilon_{1D}=2\log_e(2)=1.3862\hspace{.1cm}9436\hspace{.1cm}1119\hspace{.1cm}8906\hspace{.1cm}1883\ldots
	\label{ln2}
\end{equation}	
The analytic result in Eq.~(10) provides a convenient reference against which to test the axial constructions. Although the alternating one-dimensional sum is convergent, its ordinary partial sums approach the limit slowly. In the calculations below, each RU-$p$ construction is centered symmetrically about the origin and repeated along the chain. The charge member coincident with the reference ion is omitted from the potential sum.\\\\
Table\,\ref{tab:log2} compares the convergence of the alternating series, AltH, with the RU-5 and RU-13, where $\mathbf{n}$ is the summation radius, or shell number, in of nearest-neighbor units along the positive x-axis.The speed advantage of RU-13 is evident.\\\\
Fig.\:\ref{1D_speeds} compares the same schemes in terms of error rates. The intermediate RU-9 construction is included to further emphasize the systematic improvement obtained with increasing multipole order. The progressively steeper error-decay slopes are manifestations of increasingly faster potential decay with distance.
\begin{table}[htbp]
	\centering
	\caption{NaCl $2log(2)$ approximations, comparing convergence of the alternating harmonic series, RU-5, and RU-13. RU-13 reaches near machine precision of 14 decimal places at a very short radius. Underlining marks the first incorrect digits.}
	\begin{tabular}{clll}
		\toprule[0.2em]
		Shells & \multicolumn{3}{c}{Method}  \\
		\cmidrule{2-4}
		n & Alt-Harmonic & RU-5  & RU-13 \\
		\midrule
		4     & 1.\underline{6}  & 1.38\underline{7} & \multicolumn{1}{l}{1.38\underline{3}} \\
		10    & 1.\underline{5}  & 1.386\underline{3} & \multicolumn{1}{l}{1.3862943\underline{7}} \\
		20    & 1.3\underline{7}  & 1.38629\underline{8} & \multicolumn{1}{l}{1.3862943611\underline{2}} \\
		40    & 1.3\underline{5}  & 1.386294\underline{6} & \multicolumn{1}{l}{1.38629436111989\underline{}} \\
		100    & 1.3\underline{7}  & 1.38629436\underline{8} &  \\
		1000   & 1.38\underline{5}  & 1.3862943611\underline{2} &  \\
		\bottomrule[0.1em]
	\end{tabular}%
	\label{tab:log2}%
\end{table}
~\\\\
\FloatBarrier
\clearpage
\textit{\textbf{\underline{3D NaCl}}}\\\\
A 3D lattice is constructed in same manner of accumulating the RUs symmetrically about the reference origin.\\Figure.\,\ref{3D_speeds} compares the convergence speeds with the units RU-5, RU-7, R-9 and RU-13 in terms of their truncation errors. The summation geometry is cubical shells following the Bravais vectors. It is seen that the truncation errors decrease rapidly with shell number, with higher-order RUs producing progressively steeper convergence slopes. The RU-13 construction attains 13-decimal-places accuracy within relatively modest summation radius, approaching the practical limit of double-precision arithmetic.\\\\
Because of its efficient convergence, from this point only RU-13 will be used for discussions following.\\
\\\\
\textit{\underline{Growth Geometry Effects}}\\\\
The transition from one to three dimensions represents more than just a quantitative increase in computational effort. Whereas the one-dimensional alternating ionic chain converges naturally, the three-dimensional Coulomb lattice exhibits the classical problem of conditional convergence. Direct summation of the bare ionic charges does not generally yield a unique result, as the limiting value may depend on the geometry by which the summation domain is expanded.\\

For three-dimensional lattices, two natural growth geometries arise. The first follows the Bravais vectors through nested iteration over the Cartesian coordinates, producing a cubical growth domain. The second employs spherical growth, in which all ions enclosed within an expanding sphere are included in the summation. To examine the influence of growth geometry, RU-13 calculations were performed using both cubic and spherical domains. The results are compared in Table~\ref{geometry}.\\

The cubic and spherical growth schemes exhibit convergence essentially to the same limit. The result reflects the rapid \(r^{-13}\) decay of the RU-13 potential, which suppresses contributions from sufficiently distant regions independently of the shape of the summation domain. For both geometries, convergence to the accepted NaCl Madelung constant is obtained within approximately 40 nearest-neighbor shells. Beyond this distance, roundoff error of double-precision arithmetic is comparable to the residual truncation error. As a result, enlarging the summation domain provides little further gain, and the attainable accuracy is effectively limited to approximately 13-14 decimal places.\\

The convergence trajectories for the cubic and spherical domains were found to be nearly indistinguishable. Because their differences fall below the graphical resolution of a page-sized figure over most of the convergence range, the results are presented directly in tabular form.
 \begin{table}[htpb]
	\centering
	\caption{3D NaCl Madelung constant with RU-13, \textbf{cubical} vs \textbf{spherical} growth, as a function of multiple of nearest neighbor distance. Both methods attain 13 decimal place accuracy at $n=40$. Underlined last digits mark deviation from accepted value.}
	\begin{tabular}{cll}
			\toprule[0.2em]
			Shells & \multicolumn{2}{c}{Growth Geometry}\\
			\cmidrule{2-3}
			n & Cubic & Spherical \\
			\midrule
			3     & 1.\underline{6}   & 1.\underline{5} \\
			7     & 1.747564\underline{9} & 1.7475\underline{3} \\
			10    & 1.74756459\underline{7} & 1.7475645\underline{3} \\
			20    & 1.74756459463\underline{6} & 1.7475645946\underline{6} \\
			30    & 1.747564594633\underline{2} & 1.747564594633\underline{6} \\
			40    & 1.7475645946331\underline{7} & 1.74756459463318 \\
			\midrule
			\textit{Burrows}\cite{Burrows}   & 1.7475645946331821 & \\
			\bottomrule[0.2em]    
			\end{tabular}%
	\label{geometry}%
\end{table}~\\
\\
\textit{\textbf{Planar (2D) NaCl,\\ Center and Edge Madelung Constants}}\\\\
We next apply the same construction to a single-plane NaCl lattice. The RU-13 units are oriented parallel to the \(x\)-axis and accumulated about the reference ion at the origin as the summation domain is expanded in a square growth geometry. Potentials are then evaluated and accumulated, with any charge at the origin excluded from the sum. The edge-of-plane case is obtained by removing one half-plane, for example by excluding sites on the positive-\(y\) side. The results are listed in Table~\ref{tab:Planar}.\\

\begin{figure}[htbp]
	\centering
	\begin{subfigure}[b]{0.42\textwidth}
		\centering
		\includegraphics[width=\textwidth]{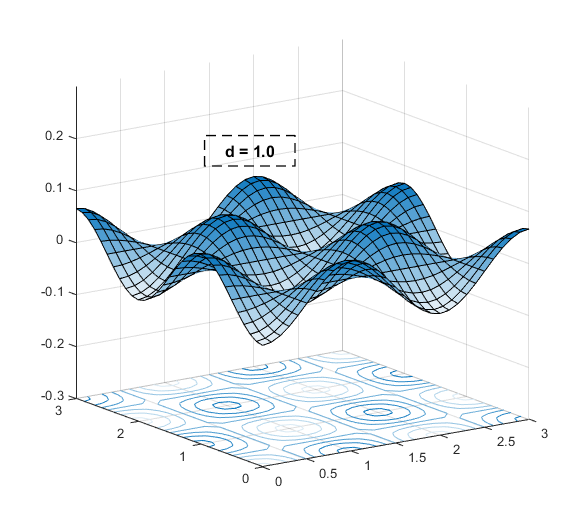}
		\caption{d = 1-unit above cleavage}
		\label{fig:image1}
	\end{subfigure}
	\hfill
	\begin{subfigure}[b]{0.50\textwidth}
		\centering
		\includegraphics[width=\textwidth]{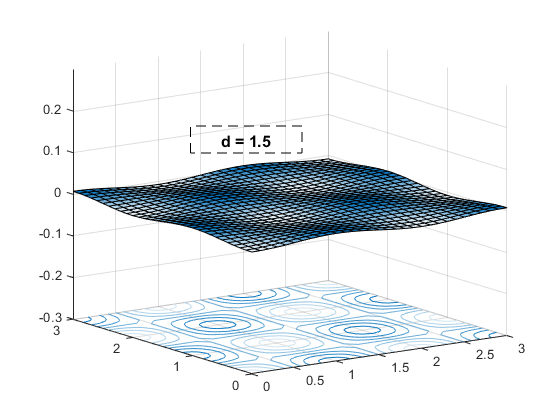}
		\caption{d = 1.5 units above cleavage}
		\label{fig:image2}
	\end{subfigure}
	\captionof{mat}{Potential plot on a $3\times3$ unit cells square off-lattice plane at different heights d above an NaCl (001) cleavage plane. One unit is Na-Cl spacing. Note how potential plot flattens to near zero at $d=1.5$, losing 90\% of its amplitude from $d=1$.}
	\label{fig:Off-lattice}
\end{figure}

\begin{figure}[htbp]
	\centering
	\includegraphics[width=.55\linewidth]{"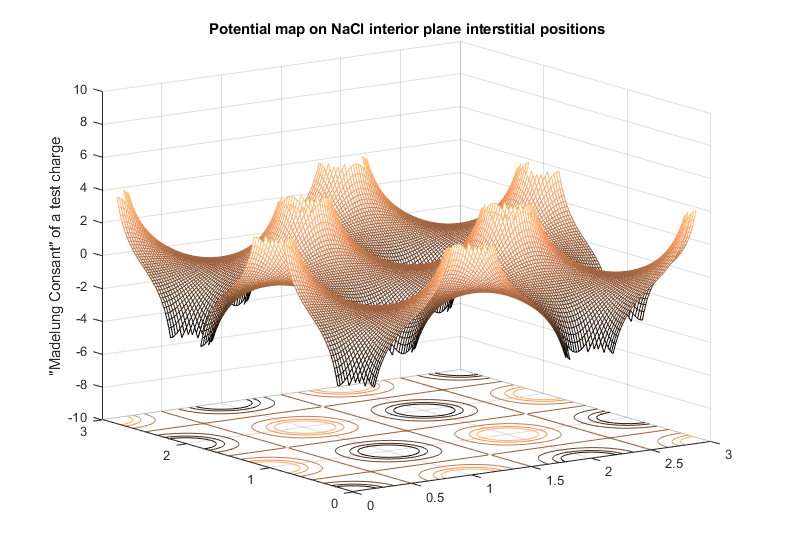"}
	\captionof{mat}{Potential plot on an interior (001) NaCl plane interstitial positions. Potential goes to $\pm\infty$ at lattice ion sites.}
	\label{In-lattice}
\end{figure}

\begin{figure}[htbp]
	\centering
	\includegraphics[width=.25\linewidth]{"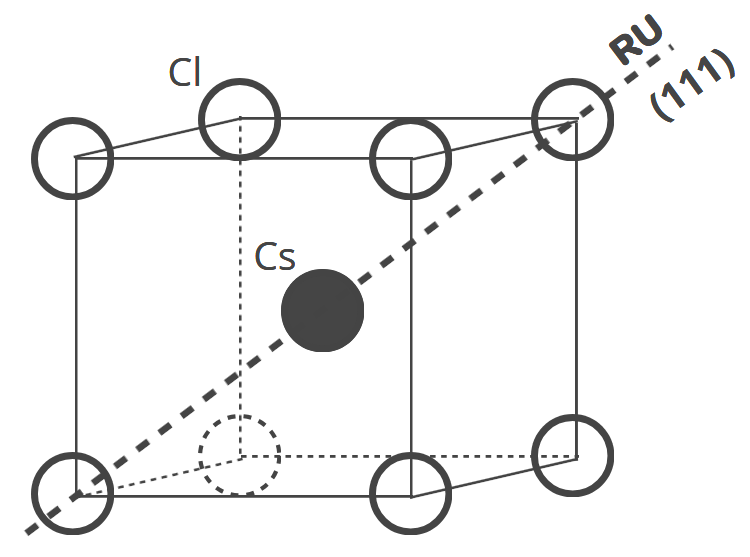"}
	\captionof{mat}{CsCl unit cell, RU-13 orientation.}
	\label{CsCl_cell_RU}
\end{figure}

For \(n=30\), the planar-center Madelung constant agrees with the value reported by Burrows~\cite{Burrows} to all digits shown. We have not found corresponding published values for the planar edge. The edge value, however, provides a useful internal check: by symmetry, it should equal one half of the planar-center value plus one half of the one-dimensional lattice value from Table~\ref{tab:log2}, namely
\[\mathrm{Edge\; MC}=(1/2)\times(1.615542626713 + 1.386294361119) = \mathbf{1.500918493916}\]
which agrees to all decimal places with the final direct sum in Table\,\ref{tab:Planar}.\\
\begin{table}[htbp]
	\centering
	\caption{Planar NaCl center and edge Madelung constants at $n$-multiples of nearest neighbor distance from reference ion. Underlined digits mark departure from Burrows's.}
		\begin{tabular}{cll}
			\toprule[0.2em]
			Shells & \multicolumn{2}{c}{Planar Madelung Constant}  \\
			\cmidrule{2-3}
			n & Center  & Edge \\
			\midrule
			10     & 1.615542\underline{4}  & 1.500918518\\
			20    & 1.6155426267\underline{2}  & 1.500918493925\\
			30    & 1.615542626713  & 1.500918493916\\
			\midrule
			\textit{Burrows}\cite{Burrows} & 1.615542626713&(None in literature)  \\
			\bottomrule[0.1em]
		\end{tabular}%
		\label{tab:Planar}%
\end{table}~\\\\
%
%
\textit{\textbf{3D NaCl Surface, and Edge Madelung Constants\\Growth Geometry}}\\\\
\textit{\underline{Surface and Edge MC}}\\\\
Surfaces and edges provide a simple test of the method under broken lattice symmetry. Table~\ref{tab:surfedge} gives the direct sums for the surface and edge Madelung constants of three-dimensional NaCl. The surface value was obtained by removing one half-space, such as the negative-(y) or negative-(z) side, and the edge value by removing both. Since the RU-13 units are aligned along the (x)-axis, the (x) direction was not used for these exclusions; doing so would truncate the RU-13 units and leave low-order multipole fragments at the boundary, slowing convergence. The table also lists values obtained from the corresponding symmetry-based formulas. If we define\\
\begin{itemize}
	\item $M1=1.386294361119890$; 1D NaCl (line) MC (2ln(2))
	\item $M2=1.615542626713$;\quad2D NaCl (planar) MC
	\item $M3=1.74756459463318$;\quad3D NaCl (bulk) MC
	\item $M_{surf}=$\quad surface MC of 3D NaCl
	\item $M_{edge}=$\quad edge MC of 3D NaCl
\end{itemize}
then
\begin{equation}
	\begin{split}
		M_{surf} &= \frac{1}{2}(M3 + M2)\\
		M_{edge} &= \frac{1}{4}M3+\frac{1}{2}(M2+M1)
	\end{split}
	\label{eqn:surfedge}
\end{equation}
%
\begin{table}[htpb]
	\centering
	\caption{3D NaCl Surface and Edge Madelung Constants. Summation radius is 40 shells.}
	\begin{tabular}{lll}
		\toprule[0.2em]
		& \multicolumn{2}{c}{3D Madelung Constants}\\
		\cmidrule{2-3}
		& Surface & Edge \\
		\midrule
		Direct sum & 1.6815536106730   &  1.5912360522947 \\
		Formulas    & 1.6815536106730 & 1.5912360522947 \\
		\midrule
		\textit{Baker}\cite{Baker}    & 1.68155361067 & 1.59123605229		 \\
		\bottomrule[0.2em]    
	\end{tabular}%
	\label{tab:surfedge}%
\end{table}
These agree to all 12 digits with Baker's\cite{Baker} \;`true values' table column.\\\\
\textit{\textbf{4D, 5D, and 6D Hyperdimensional NaCl}}\\\\
Although dimensions higher than three are not directly relevant to physical ionic crystals, hypercubic NaCl lattices provide a useful numerical testbed. The construction of the axial repeating units is independent of dimensionality, and the convergence criterion depends only on the competition between the far-field decay rate of the repeating unit and the growth of lattice sites with distance. Calculations in four through six dimensions therefore serve as a stringent validation of the method beyond the physically relevant three-dimensional case.
Programmatically, going from 3D to higher dimensions involves simply adding the requisite levels to the nested iteration on coordinates, e.g., iterating on [x y z] of 3D-cube to [x y z w] of 4D-hypercube, and so on. By default the RU-13s are oriented parallel to the x-axis.\\\\
Table \ref{tab:hypercubes} lists some results. Agreement with available literature values is good, but noticeably degrades at higher dimensions where the number of RUs accumulates at increasing rates and partially offsets the potential decay rate.\\\\
\begin{table}[htbp]
	\centering
	\caption{Madelung Constants for hypercubic NaCl of 4, 5, and 6 dimensions. Summation radius is 30 shells. For the same crystallite size along one axis, accuracy decreases with higher dimensionality.}
	\begin{tabular}{clll}
		\toprule[0.2em]
		& \multicolumn{3}{c}{Dimensionality}  \\
		\cmidrule{2-4}
		& 4D & 5D  & 6D \\
		\midrule
		This work     & 1.839399084036  & 1.9093378158 & \multicolumn{1}{l}{1.96555704} \\
		\midrule
		\textit{Burrows}\cite{Burrows}    &  1.83939908404504  & 1.90933781561876 & \multicolumn{1}{l}{1.96555703900907} \\
		\bottomrule[0.1em]
	\end{tabular}%
	\label{tab:hypercubes}%
\end{table}~
Although the RU-13 construction is formally sufficient for
absolute convergence through 12D, the required summation
volumes become rapidly prohibitive.  For this reason, the present numerical
tests are limited to dimensions up to \(D=6\), where the calculation remains
tractable.\\\\
\textit{\textbf{Potential distributions on exterior and interior planes of NaCl.}}\\\\
Potentials at arbitrary off-lattice sites are just as easily obtained as for lattice ion sites. Figure (\ref{fig:Off-lattice}) illustrates maps of potentials on a plane located parallel to a (001) cleavage plane of NaCl at different elevations. Contours projected on the X-Y plane indicate relative positions of Na, Cl on the cleavage plane, coinciding vertically with peaks and valleys of the potential surface.The z-axis represents the Madelung potential of a unit positive test charge at each point on the surface plot. The peak-to-peak amplitude diminished 90\% going from 1-unit elevation to 1.5 units.\\\\
Off-lattice potentials are evaluated with the same procedure used for lattice-ion sites. Figure (\ref{fig:Off-lattice}) shows potential maps on planes parallel to a NaCl (001) cleavage plane at several elevations above the surface. Contours projected onto the (x)-(y) plane mark the relative positions of Na and Cl ions in the cleavage plane and coincide with the extrema of the potential surface. The vertical axis gives the Madelung potential of a unit positive test charge at each point in the plane. Increasing the elevation from one to (1.5) nearest-neighbor units reduces the peak-to-peak amplitude by approximately 90\%.
Similar maps can be made for the interior lattice spaces with equal facility, as shown on Figure (\ref{In-lattice}).\\
%
~\\
\FloatBarrier
\textit{\textbf{CsCl 3D Madelung Constant}}\\
The CsCl unit cell can be represented as a cube two nearest-neighbor spacings wide, with Cl ions at the corners and a Cs ion at the body center, as shown in (\ref{CsCl_cell_RU}). Along the \{111\} body diagonal, the charges form an alternating, equally spaced Cs--Cl chain. The lattice can therefore be generated from an RU-13 axial multipole oriented along this body diagonal, with charge spacing $\sqrt{3}$. For the summation, the Cs ion is placed at the origin and the lattice is expanded outward about it. The results are given in Table~(\ref{CsCl}).
~\\
\begin{table}[htbp]
	\centering
	\caption{Convergence of the CsCl Madelung constant obtained using RU-13 axial multipoles. Distances are measured in units of the Cs--Cl nearest-neighbor spacing. Underlined digits indicate the first deviation from the accepted value.}
	\begin{tabular}{cl}
		\toprule[0.2em]
		Shells & \\
		n & Madelung Constant\\
		\midrule
		5     & 1.762\underline{9}\\
		10    & 1.76267477\underline{2}\\
		20    & 1.762674773070\underline{8}\\
		30    & 1.7626747730709\underline{4}\\
		\midrule
		\textit{Sakamoto}\cite{Sakamoto}   & 1.7626747730709883 \\
		\bottomrule[0.2em]    
	\end{tabular}%
	\label{CsCl}%
\end{table}~\\
%
\section*{Conclusions}
Axial multipoles, used here as a basis for lattice construction, provide an absolutely convergent real-space method for evaluating Madelung potentials, including Madelung constants at lattice sites. By eliminating low-order multipole moments, the resulting real-space sum is absolutely convergent, so that its limiting value is independent of the summation geometry.

The approach was demonstrated for NaCl lattices in one through six dimensions. Using the RU-13 construction, the three-dimensional NaCl Madelung constant was obtained to 13 decimal places with summation distances of only a few tens of nearest-neighbor spacings. Identical limiting values were obtained for both spherical and cubical growth geometries. The method was also applied to selected nonpolar surface and edge geometries, and to off-lattice locations, including planes above the NaCl (100) surface and representative planes within the crystal interior. It was also applied to CsCl, where the RU-13 was aligned along the unit-cell body diagonal.

The present formulation assumes that the lattice can be decomposed into neutral collinear charge arrays. It therefore does not directly apply to structures with intrinsically off-axis charge distributions, such as nitrates, carbonates, or silicates. Polar surfaces are also excluded: truncating an axial multipole at such a boundary can leave a residual net charge, leading to divergent sums. Within these limitations, high-order axial multipoles provide a simple and efficient route to accurate, absolutely convergent real-space electrostatic summation.
\\\\
\clearpage
\bibliographystyle{apsrev4-2}  
\end{document}